\documentclass[preprint,showpacs,preprintnumbers,amsmath,amssymb]{revtex4}
\usepackage{graphicx}
\usepackage{dcolumn}
\usepackage{bm}

\newcommand\rf[1]{(\ref{eq:#1})}
\newcommand\lab[1]{\label{eq:#1}}

\newcommand\br{\begin{eqnarray}}
\newcommand\er{\end{eqnarray}}
\newcommand\be{\begin{equation}}
\newcommand\ee{\end{equation}}

\newcommand\bc{\begin{center}}
\newcommand\ec{\end{center}}

\newcommand\pa{\partial}

\newcommand{\ct}[1]{\cite{#1}}

\begin{document}
\title{Coupling Electromagnetism to Global Charge}

\author
{E. I. Guendelman \footnote{e-mail: guendel@bgu.ac.il}}
\address{Physics Department, Ben Gurion University of the Negev, Beer
Sheva 84105, Israel}

\begin{abstract}
It is shown that an alternative to the standard scalar QED is possible. In this new version there is 
only global gauge invariance as far as the charged scalar fields are concerned although local gauge invariance is kept for the electromagnetic field. The electromagnetic coupling has the form
 $j_\mu (A^{\mu} +\partial^{\mu}B)$ where $B$ is an auxiliary field and the current $j_\mu$ is $A_{\mu}$
independent so that no "sea gull terms" are introduced. As a consequence of the absence of sea gulls it is seen that no Klein paradox appears in the presence of a strong square well potential. In a model of this kind spontaneous breaking of symmetry does not lead to photon mass generation, instead the Goldstone boson becomes a massless source for the electromagnetic field.  Infrared questions concerning the theory when spontaneous symmetry breaking takes place and generalizations to global vector QED are discussed. In this framework Q-Balls and other non topological solitons that owe their existence to a global $U(1)$ symmetry can be coupled to electromagnetism and could represent multiply charged particles now in search in the LHC. Furthermore we give an example where an "Emergent" Global Scalar QED  can appear from an axion photon system in an external magnetic field. Finally, formulations of Global scalar QED that allow perturbative expansions without sea gulls are developed.

\end{abstract}


\maketitle

\date{\today}

\maketitle
\section{Introduction}
In this paper it will be shown that an alternative to the standard scalar QED is possible. In this new version there is 
only global gauge invariance as far as the charged scalar fields are concerned although local gauge invariance is kept for the electromagnetic field, we call this new model Global scalar QED. The electromagnetic coupling has the form
 $j_\mu (A^{\mu} +\partial^{\mu}B)$ where $B$ is an auxiliary field and the current $j_\mu$ is $A_{\mu}$
independent so that no "sea gull terms" are introduced. In a model of this kind spontaneous breaking of symmetry does not lead to photon mass generation, instead the Goldstone boson becomes a massless source for the electromagnetic field,  infrared questions concerning the theory when spontaneous symmetry breaking takes place and generalizations to global vector QED are discussed. 

In this framework Q-Balls \cite{Q-Balls} and other non topological solitons \cite{NTPS} that owe their existence to a global $U(1)$ symmetry can be coupled to electromagnetism and could represent multiply charged particles now in search in the LHC  \cite{Romaniouk}.

We also give an example where an "Emergent" Global Scalar QED  can appear from an axion photon system in an external magnetic field. 

Finally formulations of Global scalar QED that allow perturbative expansions involving all excitations of the theory simultaneously are developed.

\section{Conventional scalar QED and its sea gulls}
In  conventional scalar QED, we "minimally couple" a globally invariant action (under global phase transformations). To be concrete, for a complex scalar field  $\psi$ with mass, $m$ whose Lagrangian density can be represented in relativistic invariant form in the absence of interactions to electromagnetism as
\begin{equation} \label{freeComplexScalar}
\mathcal{L}=\not h^2 g^{\mu\nu}\frac{\partial\psi^{*}}{\partial x^{\mu}}\frac{\partial \psi}{\partial x^{\nu}}-m^{2}c^2\psi^{*}\psi
\end{equation}

Then, in the standard scalar QED model we introduce the electromagnetic interaction with scalar charged particles by introducing the  minimal coupling in the Lagrangian for charged particles (see Eq. \ref{freeComplexScalar}).  As we recall, minimal coupling requires that we let the momentum $p_{\mu}$ be replaced by $p_{\mu}\rightarrow p_{\mu}-eA_{\mu}$ where $p_{\mu}=-i\not h \frac{\partial}{\partial x^{\mu}}$ and where $A_{\mu}$ is the electromagnetic 4-vector whose Lagrangian is given by
\begin{equation}
\mathcal{L_{EM}}=-\frac{1}{4} F^{\mu \nu}F_{\mu \nu}
\end{equation}
with $F^{\mu \nu}=\partial^{\mu}A^{\nu}-\partial^{\nu}A^{\mu}$. We can now write the total Lagrangian after using the minimal coupling substitution into  Eq. \ref{freeComplexScalar}

\begin{equation} \label{scalarIntChargeLagran}
\mathcal{L_{T}}=g^{\mu \nu}\left[(\not h\frac{\partial}{\partial x^{\mu}}-ieA_{\mu})\psi^{*}][(\not h \frac{\partial}{\partial x^{\nu}}+ ieA_{\nu})\psi\right]-m^{2}c^2\psi^{*}\psi-\frac{1}{4} F^{\mu \nu}F_{\mu \nu}
\end{equation}

This leads to the equation of motion for the scalar field  $\psi$
\begin{equation}	\label{KG:external field}
(i \not h \frac{\partial}{\partial t} - e \phi)^{2} \psi=(\frac{c\not h}{i}\nabla - e \mathbf{A})^{2}\psi +m^{2}c^{4}\psi
\end{equation}

  This equation and the lagrangian density from which it is derived are invariant under local gauge transformations:
\begin{equation} 
\mathbf{A} \rightarrow \mathbf{A}'= \mathbf{A} + \nabla \chi \textrm{;   } \ \ \ 
\phi \rightarrow \phi ' = \phi - \frac{1}{c} \frac{\partial \chi}{\partial t} \ \ \  \textrm{with} \ \ \ \psi \rightarrow \exp{[\frac{ie\chi}{\not h c}]} \psi
\end{equation}

 Furthermore the electromagnetic field satisfies the Maxwell's equations where the electric charge density $\rho$ and the current density $\mathbf{j}(x)$ are given by (now set $c=\not h=1$).
\begin{equation} \label{KG:chargedensityandcurrent}
\rho (x)= ie(\psi^{*}\frac{\partial\psi}{\partial t}  -\psi\frac{\partial\psi^{*}}{\partial t}) -2e^2\phi \psi^{*}\psi \textrm{  and  } \mathbf{j}(x)= -ie(\psi^{*}\mathbf{\nabla}\psi  -\psi\mathbf{\nabla}\psi^{*}) -2e^2 \mathbf{A}\psi^{*}\psi
\end{equation}

There is an example, the BCS theory of superconductivity \cite{BCS},  where the effective theory in terms of 
the composite Cooper pairs retains the local gauge invariance which involves the local phase transformations of the composite scalar, however we may ask if this is a general rule, may be not.

When thinking of the electromagnetic interactions of pions, the quadratic dependence of the interactions on the potentials characterises the sea gull behaviour of standard scalar QED. As pointed out by Feynman \cite{Feynman}, it is somewhat puzzling that spinor electrodynamics  does not lead to any of such sea gulls . Considering that the microscopic description of charged pions is really the spinor electrodynamics of quarks, shouldn't we search for an effective scalar electrodynamics devoid of sea gulls?, is this possible?. In the next section we will see  that this can be achieved in global scalar QED. The Global Scalar QED could address other questions as well. like the electromagnetic  coupling of Q-balls and can "emerge" as an effective description of a system of axions and photons in an external field.

\section{Global Scalar QED, the simplest model}
There are many possible motivations for departing from the scheme implied by the minimal coupling, which leads to scalar QED. For example, if the complex scalar field is to describe a pion, since the macroscopic hadron is a very non local construction in terms of the fundamental quark fields and gluon fields as has been revealed  from both the theoretical point of view \cite{pionwf} and from the experimental point of view \cite{pionwf1} and in fact we may have several alternative candidates for the pion wave function (and any such proposal could give rise to a different effective theory),  we do not necessarily have to keep a local gauge invariance in terms of the composite scalar fields
(that would describe the hadrons), although global phase invariance must be respected. Also local gauge transformations for the photon should be mantained. Other possible use of deviating from the the minimal coupling scheme, as we will see, could be to couple Q-Ball type solitons to electromagnetism. Finally, we will give an example where an "Emergent" Global Scalar QED  can appear from an axion photon system in an external magnetic field.

We work therefore with the following lagrangian density

\begin{equation} \label{GlobalQED}
\mathcal{L}= g^{\mu\nu}\frac{\partial\psi^{*}}{\partial x^{\mu}}\frac{\partial \psi}{\partial x^{\nu}}-U(\psi^{*}\psi) -\frac{1}{4} F^{\mu \nu}F_{\mu \nu} + j_\mu (A^{\mu} +\partial^{\mu}B)
\end{equation}
 
 where
 
 \begin{equation}
 j_\mu =  ie(\psi^{*}\frac{\partial\psi}{\partial x^\mu}  -\psi\frac{\partial\psi^{*}}{\partial x^\mu})
\end{equation}

and where we have also allowed an arbitrary potential $U(\psi^{*}\psi)$ to allow for the possibility of spontaneous breaking of symmetry.
The model is separately invariant under local gauge transformations
\begin{equation} \label{GTGlobalQED}
A^\mu \rightarrow A^\mu + \partial^\mu \Lambda \textrm{;   } \ \ \ 
B \rightarrow  B - \Lambda 
\end{equation}

and the independent global phase transformations
\begin{equation}
\psi \rightarrow exp (i\chi) \psi
\end{equation} 

 The use of a gauge invariant combination $(A^{\mu} +\partial^{\mu}B)$ can be utilized for the construction of mass terms\cite{Stueckelberg}  or both mass terms and couplings to a current defined from the gradient of a scalar in the form $(A^{\mu} +\partial^{\mu}B)\partial_{\mu}A$ 
 \cite{Guendelman}, these kind of contributions will be considered in other formulations of Global Scalar QED. In the non abelian case mass terms constructed along these lines have been considered by Cornwall \cite{Cornwall}. Since the subject of this paper is electromagnetic couplings of photons and there is absolutely no evidence for a photon mass, we will disregard such type of mass terms and concentrate on the implications of the
 $(A^{\mu} +\partial^{\mu}B)j_{\mu}$ couplings. It is also interesting to point out the use of scalars instead of vector fields has been studied in \cite{Singleton}
in their general study of gauge procedure with gauge fields of various ranks.
\section{Symmetries, Conserved Charges, Constraints of the Theory Gauge fixing and remarks on the quantum theory}
 As we will see the scalar QED model has two charge conservation laws associated with it.
 We see that Maxwell's equations are satisfied with $ j_\mu$ being the source, that is 
\begin{equation}
\partial^\nu F_{\nu \mu } = j_\mu 
\end{equation}
 of course this implies $\partial^\nu \partial^\mu F_{\nu \mu } = \partial^\mu j_\mu = 0$.
 The same conclusion can be obtained from the equation of motion obtained from the variation with respect to $B$.
 This conserved quantity is a consequence of the global symmetry $B \rightarrow  B - \Lambda$ in the case
 $\Lambda = constant$, indeed the Noether current that arises from this global symmetry is exactly the source of the electromagnetic field, that is $ j_\mu $.
  
 The Noether current obtained from the independent global phase transformations
$\psi \rightarrow exp (i\chi) \psi$,  $\chi$ being a constant, is 

\begin{equation}
 J_\mu =  ie(\psi^{*}\frac{\partial\psi}{\partial x^\mu}  -\psi\frac{\partial\psi^{*}}{\partial x^\mu})
 + 2e^{2} (A_{\mu} +\partial_{\mu}B)\psi^{*}\psi
\end{equation}

Therefore 
\begin{equation}
 j^B_\mu =J_\mu - j_\mu = 2e^{2} (A_{\mu} +\partial_{\mu}B)\psi^{*}\psi
\end{equation} 

  is also conserved, that is $\partial^\mu((A_{\mu} +\partial_{\mu}B)\psi^{*}\psi) = 0$
  
  As for the local symmetries, it is well known that they are not associated to non trivial conserved charges, but rather to first class constraints, that is a set of constraints whose Poisson brackets between them vanish. Denoting the canonically conjugate momenta to $A^0$, $A^i$ and $B$  as $\pi^0 $ $\pi^i$ and  $\pi_B$ respectively, those first class constraints are  
\begin{equation}
 \pi^0 \approx 0
\end{equation} 
and 
\begin{equation}
\partial_i\pi^i + \pi_B \approx 0
\end{equation} 

These constraints have to be considered together with those related to the ones that result from gauge fixing.  One particularly interesting gauge is the $B=0$ gauge. 

Once the $B=0$ gauge an additional constraint is generated that allows to formally solve $A^0$ in terms of other dynamical variables. To see this let us consider the definition of $\pi_B$
\begin{equation}
 \pi_B = j^0 = ie(\psi^{*}\frac{\partial\psi}{\partial x^0}  -\psi\frac{\partial\psi^{*}}{\partial x^0})
\end{equation}
  
We  now express the time derivatives of the charged scalar field in terms of the canonical conjugate momenta of  $\psi$ and $\psi^{*}$ which we will denote as $\pi$ and $\pi^{*}$ 
we obtain
\begin{equation}
\frac{\partial\psi}{\partial x^0} = \pi^{*} + ie \psi(A_{0}+ \partial_{0}B)
\end{equation} 
 
 and
 \begin{equation}
\frac{\partial\psi^{*}}{\partial x^0} = \pi - ie \psi^{*}(A_{0}+ \partial_{0}B)
\end{equation}-
Finally, expressing then  $\pi_B$ in terms of $\pi$ and $\pi^{*}$ we obtain
\begin{equation}
 \pi_B = ie(\pi^{*}\psi^{*} - \pi \psi) -2e^{2}\psi^{*}\psi(A_{0}+ \partial_{0}B)
\end{equation} 
At this point we see that setting the "unitary gauge" $B=0$ allows us to solve $A_{0}$ in terms of other canonical variables, since ts own canonically conjugate momentum vanishes, obtaining
\begin{equation}
A_{0} = \frac{ie(\pi^{*}\psi^{*} - \pi \psi) - \pi_B}{2e^{2}\psi^{*}\psi} = \frac{ie(\pi^{*}\psi^{*} - \pi \psi)+ \partial_i\pi^i }{2e^{2}\psi^{*}\psi}
\end{equation} 
 
Formally  the above equation takes care of the all the constraints in this gauge, the zero component of the gauge field is eliminated and under quantization the spatial components of the gauge fields and their respective canonically conjugate momenta satisfy canonical equal time commutations. The model in the gauge $B=0$ does not reduce to any known model.

One should notice however that the above solution for $A_{0}$ and therefore the whole idea of using the $B=0$ gauge makes perfect sense only when $\psi^{*}\psi$ has a non zero expectation value, otherwise will frequently will have to face the problem of dividing by zero when using such solution for $A_{0}$.

One should conclude that if there is no spontaneous symmetry breaking of the global symmetry
 $\psi \rightarrow exp(i\theta)\psi$, that is, if $\psi^{*}\psi$ has a zero expectation value, then such a gauge is not possible. Attempting to use such a gauge would be like trying to use in standard scalar QED the phase of the complex scalar field when dealing with the unbroken phase of the theory, which we know is inconsistent since at $\psi^{*}\psi = 0$ such a phase is not defined, if there is spontaneous symmetry breaking of the global symmetry $\psi \rightarrow exp(i\theta)\psi$, that is, if $\psi^{*}\psi$ has a non zero expectation value, we know that unitary gauge is possible.
 
 Furthermore, from the point of perturbation theory, in the unbroken theory, it is a bad idea to use $B$ in the gauge fixing, since
  $B$ appears in the interaction and not in the free part of the lagrangian. A gauge like Coulomb gauge would be preferable, but then the field  $B$ is not determined to zeroth order in perturbation theory.
 The theory as it is appears perfectly well defined, but perturbation theory in this simple formulation seems problematic in the unbroken phase. We will see later in the paper that it is possible to have slightly more elaborate models that allow a well defined perturbation theory in the unbroken phase.
  
Also, since the  $B$ field does not appear in the free field theory, the free photons have as usual only  transverse polarizations, as in normal electrodynamics.

Finally, some preliminary observations concerning the quantum theory and radiative corrections, for example thinking of the theory in the broken symmetry case where a pertubation theory is not problematic, should be made: first, the model can be renormalizable if the potential   $U(\psi^{*}\psi)$ is taken to be of the form  $U(\psi^{*}\psi) =  m^{2} (\psi^{*}\psi)  + \lambda (\psi^{*}\psi)^{2}$. Second, under renormalization, in order the keep the photon massless, there could be the need to introduce 
counter terms of the form $\delta \mu^{2}(A^{\mu} +\partial^{\mu}B)(A_{\mu} +\partial_{\mu}B)$. This should require further investigations.Similar discussion should be given for formulations of global scalar QED  that allow a perturbative expansion in the unbroken phase of the theory.

\section{Degrees of Freedom of the Free Photon}  
It is very important to notice that for the free photons, far from the source ($\psi$) and arranging that the physical mass is zero, we obtain exactly the same equations as those of standard  Maxwell theory, where the $B$ field does not appear at all, in this case the  $B=0$ gauge does not fix a gauge for the vector field, since it cannot be used in the regions where  ($\psi =0$).

As a result of this, free photons are massless and with only two independent polarizations, since the theory in the $e \rightarrow 0$ is exactly standard Maxwell electrodynamics.

\section{No Klein Paradox as a consequence of the absence of sea gulls in global scalar QED}
An interesting difference between standard scalar QED and global scalar QED appears in the case of strong fields. We will see that no Klein Paradox is present  in global scalar QED as a consequence of the absence of sea gulls.
Consider the global scalar QED equations  with an external electromagnetic field potential step-function: $e(A_{0}+ \partial_{0}B)\equiv V(x); \ eA_{i}+ e\partial_{i}B=0$.
\begin{eqnarray}
V(x)=\left\{ \begin{array}{ll} 0 \ \ \ \ \textrm{for} \ \ \ x<0  \nonumber  \\
V_{0} \ \ \ \textrm{for} \ \ x>0 \ \ \  \end{array}\right.
\end{eqnarray}

The Global QED equation in the presence of this potential is ($\not h=1,c=1$)
\begin{equation} \label{eq:free KG}
-\frac{\partial^{2}\psi}{\partial t^{2}}+\mathbf{\nabla}^{2}\psi-m^{2}\psi=0
\end{equation}
for $x<0$.
\begin{equation} \label{eq:KG with V}
-\frac{\partial^{2} \psi}{\partial t^{2}}+ 2iV_{0}\frac{\partial \psi}{\partial t}+\mathbf{\nabla}^{2}\psi-m^{2}\psi=0
\end{equation}
for $x>0$.
To solve the equation with this potential, we try solutions of the form:
\begin{eqnarray} \label{eq:KG paradox}
\psi_{<} & \equiv &\psi=e^{-iEt}[e^{ipx}+Re^{-ipx}] \ \ \textrm{for}\ \ x<0 \nonumber \\
\psi_{>} &\equiv & \psi=Te^{-iEt}e^{ip'x}\ \ \textrm{for}\ \ x>0
\end{eqnarray}
where $\psi_{<}$ represents a wave like solution for the Klein-Gordon field for $x<0$ and $\psi_{>}$ represent the field for wave like solution for $x>0$.  $R$ is the amplitude of that part of wave that is reflected wave while T is that part that is transmitted.  We substitute $\psi_{<}$ and $\psi_{>}$,  Eq. \ref{eq:KG paradox} into Eqs. \ref{eq:free KG} and \ref{eq:KG with V} respectively.  We thus find
\begin{equation}
E^{2}-p^{2}-m^{2}=0 \rightarrow E=+\sqrt{p^{2}+m^{2}}  \ \textrm{for} \ \ x<0
\end{equation}
since for incident wave for $x<0$ we chose the positive sign in the square root as our boundary condition.
and

\begin{equation}
E^{2}-2EV_{0}-p'^{2}-m^{2}=0 \rightarrow p'=\pm\sqrt{E(E-2V_{0})-m^{2}} \ \ \textrm{for} \ \ x>0
\end{equation}

We see here that from a certain positive value of $V_{0}$, $V_{0crit}= (E^{2}-m^{2})/2E$ and higher,$p'$
becomes imaginary and therefore there is no transmited wave for large  values of $V_{0}$, totally opposite to the behaviour of standard scalar QED, where for large enough barrier a transmitted wave is restored once again, leading to the "Klein paradox", the transmitted wave is interpreted there as pair creation process, no such process appears in global scalar QED.

Other processes where the absence of sea gulls could be important should be for example the Compton effect in global scalar QED. This will be studied in a future publication.
 
\section{Behaviour under Spontaneous breaking of symmetry, new couplings of Goldstone Bosons to Electromagnetism  and associated infrared problems}
The absence of quadratic terms in the vector potential implies that no mass generation for the photon takes place.
Furthermore the Goldstone boson that results from this s.s.b. ,writing 
$\psi= \rho exp(i\theta) $ , where $\rho$ is real and positive, we obtain that
the phase of the $\psi$
field, is not eaten, it remains in the theory, in fact it couples derivatively to  
$(A_{\mu} +\partial_{\mu}B)$, like the $A$ field studied in \cite{Guendelman}
and it produces a gradient type charge. In fact under s.s.b. regarding $\rho$ as a 
constant, $j^\mu = 2e\rho^2\partial^\mu\theta$ the coupling 
$(A_{\mu} +\partial_{\mu}B)j^\mu$ implies the coupling of 
$(A_{\mu} +\partial_{\mu}B)$ to a gradient current, as discussed in \cite{Guendelman}.

It should be pointed out that this type of gradient current  $j^\mu = 2e\rho^2\partial^\mu\theta$ for 
 $\rho = constant$ generates an infrared problem, since the  $\theta$ field now represents a massless field, which instead of being eaten becomes a source of electromagnetism. The normal way of solving for  the electromagnetic field, using the Green's function method does not work straightforwardly, since the source now in Fourier space has support only in the light-cone and the Green's function has a pole like behaviour at the light-cone as well, so we encounter an undefined product of distributions. This is very similar to the solution of a forced harmonic oscillator when the external force has exactly the same frequency to that of the oscillator, that is the resonant case.  

To resolve this problem, we note first that considering $ F_{\nu \mu }$ as an antisymmetric tensor field (without at first considering whether this field derives from a four vector potential), then a solution of the equation 
$\partial^\nu F_{\nu \mu } = j_\mu $  is \footnote{I want to thank R. Tabensky for pointing this to me}
\begin{equation}
 F_{\nu \mu } = \int_0^1 d\lambda \lambda^2 (x_\nu j_\mu(\lambda x) -  x_\mu j_\nu(\lambda x))
\end{equation}
For a generic current the above $F_{\nu \mu }$ does not derive from a potential, however if the current is the gradient of a scalar field, the above $F_{\nu \mu }$  derives from a potential and provides a solution of the problem, where the Green's function method fails. Notice that the similarity with the the resonant case of the forced harmonic oscillator is very close, there the solution is of the form of an oscillating
function times time and in the above solution we see the similar  $x_\nu$ dependence appearing.

The resulting gauge potentials displays also a linear dependence on $x_\nu$, which is interesting, since the central issue in the confinement problem for example is how to obtain potentials with linear dependence on the  coordinates, although it is not clear how the very specific solution studied here is relevant to the confinement problem. 

It is interesting to note that the $B$ equation is very simple to first order in the spontaneously broken phase, indeed, then $\partial^\mu((A_{\mu} +\partial_{\mu}B)\psi^{*}\psi) = 0$ , taking 
$\psi^{*}\psi = constant$, reduces to $\partial^\mu A_{\mu} + \Box B = 0$, which is formally solved as
$B= -\frac{1}{\Box}\partial^\mu A_{\mu}$.

\section{Global Vector QED}
In this case we consider a complex vector field $W_\mu$ and consider the action 
\begin{equation} \label{GlobalVQED}
\mathcal{L}= -\frac{1}{4} g^{\mu\nu}g^{\alpha\beta}G_{\mu\alpha}G^*_{\nu\beta} -\frac{1}{4} F^{\mu \nu}F_{\mu \nu} + j_\mu (A^{\mu} +\partial^{\mu}B) + M^2 W_\mu W^{*\mu}
\end{equation}
 
with $G^{\mu \nu}=\partial^{\mu}W^{\nu}-\partial^{\nu}W^{\mu}$ and where
 
 \begin{equation}
 j_\mu =  ie(W^{*\alpha}G_{\alpha \mu}  -W^\alpha G^{*}_{\alpha \mu})
\end{equation}
This model displays global phase invariance for the complex vector field $W_\mu$ and local gauge invariance for the photon and $B$ fields (\ref{GlobalQED}), as was the case of global scalar QED. Once again, no sea gull terms are present here.

In regard to this global vector QED model, there are some resemblances  with the model obtained by gauging the dual symmetry between 
electric and magnetic quantities of Maxwell's equations by  Kato and  Singleton  \cite{KSingl} where they also end up with a complex vector field as we do here.
They did not include the real vector field and did not include a mass term for the complex vector. 

\section{ Q Balls and other  global $U(1)$ solitons as electromagnetically charged Particles}
An interesting situation could present itself when considering solitons  as in the case of Q-Balls \cite{Q-Balls}
or other non topological solitons \cite{NTPS}, that depend on the existence of a $U(1)$ symmetry.

These solitons have been found using actions like that used in Global scalar QED for the case $e$=0.
The idea is minimizing the energy under the constraint that the charge of the system is given.
This leads us to time dependent configurations with time dependence of the form
 \begin{equation} \label{timedependence}
\psi(r, t) =  \rho(r) exp(i\omega t)
\end{equation}

We see that if there was a local gauge transformation that involve a local phase transformation of the complex scalar field $\psi$, then the phase of $\psi$ is a totally unphysical quantity and the above eq.
\ref{timedependence}
 becomes totally meaningless. That is not the case in global QED, for which \ref{timedependence}
is meaningful.

Furthermore, the standard Q-Balls hold in the limit $e \rightarrow 0$ and also the small $e$ case can be treated in perturbation theory, The introduction of a non zero $e$ tends to destabilize the soliton as a consequence of the Coulomb repulsion that appears from the Q-ball having an electric charge. This effect is small for the case of  small $e$, so we know there must be a range of parameters for which electrically coupled Q-Ball solitons exist.

\section{"Emergent" scalar QED from a system of photons and axions in an external magnetic field}
The possible existence of a light pseudo scalar particle is a very interesting possibility. For example,
the axion \cite{Peccei}, \cite{weinberg}, \cite{Wilczek}  which was introduced in order to solve the strong CP problem has since then also been postulated as a candidate for the dark matter.
A great number of ideas and experiments for the search this particle have been proposed
\cite{Goldman}, \cite{Review}.

Here we are going to focus on a particular feature of the axion field $\phi$: its coupling to the photon 
through an interaction term of the form $g \phi \epsilon^{\mu \nu \alpha \beta}F_{\mu \nu} F_{\alpha \beta}$.
In fact, a coupling of this sort is natural for any pseudoscalar interacting with
electromagnetism, as is the case of the neutral pion coupling to photons (which, as a consequence of this interaction decays into two photons).

It was recognized by Sikivie that axion detection exploiting axion to photon
conversion in a magnetic field was a possibility \cite{Sikivie}.

A way to explore for observable consequences of the coupling of a light scalar 
to the photon in this way is to subject a beam of photons to a very strong magnetic field. This affects the optical properties of light which could lead to testable consequences \cite{PVLAS}. Also, a magnetic field in the early universe can lead to interesting photon-axion conversion effects \cite{Yanagida} and in the laboratory photon-axion conversion effects could be
responsible for the "light shining through a wall phenomena ", which are obtained by first
producing axions out of photons in a strong magnetic field region, then subjecting the mixed beam of photons
and axions to an absorbing wall for photons, but almost totally transparent to axions due to their weak 
interacting properties which can then
go through behind this "wall", applying then another magnetic field one can recover once again some photons 
from the produced axions \cite{LSTW}, \cite{Rabadan}.

In this section we will consider how an "Emergent" scalar QED from a system of photons and axions in an external magnetic field. Such analysis was considered in \cite{duality} and in \cite{splt}, where a "scalar QED analogy" was recognized. As we will discuss here, although the system of photons and axions in an external magnetic field does indeed have features that resemble scalar QED, the more close correspondence is with Global Scalar QED.

The action principle describing the relevant light pseudoscalar coupling to the photon is

\begin{equation}
\label{axion photon ac }
S =  \int d^{4}x 
\left[ -\frac{1}{4}F^{\mu\nu}F_{\mu\nu} + \frac{1}{2}\partial_{\mu}\phi \partial^{\mu}\phi - 
\frac{1}{2}m^{2}\phi^{2} - 
\frac{g}{8} \phi \epsilon^{\mu \nu \alpha \beta}F_{\mu \nu} F_{\alpha \beta}\right].
\end{equation}

We now specialize to the case where we consider an electromagnetic field with propagation along the $y$ and $z$ directions and where a strong magnetic field pointing in the $x$-direction is present. This field may have an arbitrary space dependence in $y$ and $z$, but it is assumed to be time independent.

For the small perturbations, we consider only small quadratic terms in the action for the axion and the electromagnetic fields, considering a static magnetic field pointing in the $x$ direction having an arbitrary $y$ and $z$ dependence and specializing to $y$ and $z$ dependent electromagnetic field perturbations and axion fields. This means that the interaction between the background field , the axion and photon fields reduces to 
 
\begin{equation}
\label{axion photon int }
S_I =  - \int d^{4}x 
\left[ \beta \phi E_x \right],
\end{equation}

where $\beta = gB(y,z) $. Choosing the temporal gauge for the photon excitations and considering only the $x$-polarization for the electromagnetic waves (since only this polarization couples to the axion) we get the following 2+1 effective dimensional action (A being the x-polarization of the photon, so that $E_x = -\pa_{t}A$)

\begin{equation}
\label{2 action}
S_2 =  \int dydzdt 
\left[  \frac{1}{2}\partial_{\mu}A \pa^{\mu}A+ \frac{1}{2}\partial_{\mu}\phi \pa^{\mu}\phi - 
\frac{1}{2}m^{2}\phi^{2} + \beta \phi \pa_{t} A \right].
\end{equation}

Since we consider only $A=A(t,y, z)$, $\phi =\phi(t,y,z)$, we have avoided the integration over $x$. For
the same reason $\mu$ runs over $t$, $y$ and $z$  only . This leads to the equations

\begin{equation}
\label{eq. ax}
\pa_{\mu}\pa^{\mu}\phi + m^{2}\phi =  \beta \pa_{t} A
\end{equation}

and

\begin{equation}
\label{ eq. photon}
\pa_{\mu} \pa^{\mu}A = - \beta \pa_{t}\phi.
\end{equation}

As is well known, when choosing the temporal gauge the action principle cannot reproduce the Gauss 
constraint (here with a charge density obtained from the axion photon coupling) and has
to be imposed as a complementary condition. However this 
constraint is automatically satisfied here just because of the type of dynamical reduction
employed and does not need to be considered  any more.

Without assuming any particular $y$ and $z$-dependence for $\beta$, but still insisting that 
it will be static, we see that in the case $m=0$, we discover a continuous axion 
photon duality symmetry (these results were discussed previously in the 1+1 dimensional case, where only $z$ dependence was considered in \cite{duality} and generalized for the case of two spatial dimensions in \cite{splt}), since

\begin{enumerate}

  \item  The kinetic terms of the photon and
axion allow for a rotational $O(2)$ symmetry in the axion-photon field space.
  \item The interaction term, after dropping  a total time derivative, can also be expressed in 
an $O(2)$ symmetric way as follows:

\end{enumerate}

\begin{equation}
\label{axion photon int2}
S_I =  \frac{1}{2} \int dydzdt 
\beta \left[ \phi \pa_{t} A - A \pa_{t}\phi \right].
\end{equation}

It is easy to see that after introducing an appropriate complex field $\phi$, this coupling is exactly of the global scalar QED form. 
The $U(1)$ axion photon symmetry is (in the infinitesimal limit)

\begin{equation}
\label{axion photon symm}
\delta A = \epsilon \phi, \delta \phi = - \epsilon A,
\end{equation}

where $\epsilon$ is a small number. Using Noether`s theorem, this leads to the 
conserved current $j_{\mu}$, with components given by

\be
j^{N}_{0} = A \pa_{t}\phi - \phi \pa_{t} A - \frac{\beta}{2}(A^{2} + \phi^{2} )
\lab{axion photon density}
\ee

and 

\be
j^{N}_{i} = A \pa_{i}\phi - \phi \pa_{i} A.
\lab{axion photon current}
\ee

Here $i= y, z$ coordinates. In order to have the exact correspondence with Global scalar QED, we must define the complex field $\psi$ as
\be
\psi = \frac{1}{\sqrt{2}}(\phi + iA),
\lab{axion photon complex}
\ee
we see that 
in terms of this complex field, the Noether charge density takes the form
\be
j^{N}_{0} = i( \psi^{*}\pa_{t}\psi - \psi \pa_{t} \psi^{*}) -  \beta \psi^{*}\psi.
\lab{axion photon density complex}
\ee

which, as in Global scalar QED does not coincide with the current that enters in the interaction lagrangian, which is
\be
j_{0} = i( \psi^{*}\pa_{t}\psi - \psi \pa_{t} \psi^{*}) 
\lab{axion photon density of interaction}
\ee

We observe that the correspondence with standard scalar QED is approximate, only to first order in $\beta$, since (\ref{axion photon int2}) which represents the
interaction of the magnetic field couples with the "axion photon density" 
\rf{axion photon density of interaction}, that does not contain $\beta$ dependence.

This interaction has exactly the same form as that of the global scalar QED with an external "electric " field. In fact the magnetic field (or more precisely $\beta /2$) appears to play the role of external electric potential of Global scalar QED $e(A_{0}+ \partial_{0}B)\equiv V(x)$ that couples to the axion photon density,\rf{axion photon density of interaction} which plays the role of an electric charge density, exactly as in Global Scalar QED.

From the point of view of the axion-photon conversion experiments,
the symmetry (\ref{axion photon symm}) and its finite form, which is just
a rotation in the axion-photon space, implies a corresponding symmetry
of the axion-photon conversion amplitudes, for the limit $\omega >>m$.

In terms of the complex field, the Noether current takes the form
\be
j^{N}_{k} = i( \psi^{*}\pa_{k}\psi - \psi \pa_{k} \psi^{*}). 
\lab{axion photon current complex}
\ee

Let us introduce the charge conjugation 

\be
\psi \rightarrow \psi^{*}. 
\label{charge conjugation}
\ee

We see then, that the free part of the action  is indeed invariant under (\ref{charge conjugation}).
The $A$ and $\phi$ fields when acting on the free vacuum give rise to a photon and an axion respectivelly,
but in terms of the particles and antiparticles defined in terms of  $\psi$, we see that a photon is an antisymmetric combination of particle and antiparticle and an axion a symmetric combination, since

\be
\phi =\frac{1}{\sqrt{2}}(\psi^{*} +\psi), ~A= \frac{1}{i\sqrt{2}}(\psi - \psi^{*}),
\lab{part, antipart}
\ee

so that the axion is even under charge conjugation, while the photon is odd.
These two eigenstates of charge conjugation will propagate without mixing as long as no external magnetic field is applied.
The interaction with the extenal magnetic field is not invariant under (\ref{charge conjugation}). In fact,
under (\ref{charge conjugation}) we can see that
\be
S_I \rightarrow - S_I.
\lab{non invariance}
\ee

Therefore  these symmetric and antisymmetric combinations, corresponding to axion and photon are not going to be maintained in the presence of $B$  in the analog Global QED language, since the "analog external electric potential" breaks the symmetry between particle and antiparticle and therefore will not keep in time the symmetric or antisymmetric combinations.  In fact if the analog external electric potential is taken to be a
repulsive potential for particles, it will be an attractive potential for antiparticles, so the symmetry breaking is maximal.

Even at the classical level these two components suffer opposite forces, so under the influence of an inhomogeneous magnetic field both a photon or an axion will be decomposed through scattering into their particle and antiparticle components, each of which is scattered in a different direction, since the analog electric force is related to the gradient of the effective electric potential, i.e., the gradient of the magnetic field, times the $U(1)$ charge which is opposite for particles and antiparticles.

For this effect to have meaning, we have to work at least in a 2+1 formalism \cite{splt}, the 1+1 reduction \cite{duality},  which allows motion only in a single spacial direction is unable to produce such separation, since in order to separate particle and antiparticle components we need at least two dimensions
to obtain a final state with particles and antiparticles going in slightly different directions.

This is in a way similar to the Stern Gerlach experiment in atomic physics
\cite{Stern Gerlach}, where different spin orientations suffer a different force proportional to the gradient of the magnetic field in the direction of the spin. Here instead of spin we have that the photon is a combination of two states with different $U(1)$ charge and each of these components will suffer opposite force under the influence of the external inhomogeneous magnetic field. Notice also that since particle and antiparticles are distinguishable, there are no interference effect between the two processes.

Therefore an original beam of photons will be decomposed through scattering into two different elementary particle and antiparticle components
plus the photons that have not undergone scattering. These two beams are observable, since they have both photon components, so the observable
consequence of the axion photon coupling will be the splitting by a magnetic field of a photon beam. This effect being however an effect of first order in the axion photon coupling, unlike the ``light shining through a wall phenomena''. 

Notice that the strict scalar electrodynamics analog of the axion photon system in an external magnetic field is only with Global scalar QED, there is only similarity with standard scalar QED, which requires disregarding sea gull terms, i.e., terms quadratic in the external gauge field potentials. Things that depend on these quadratic terms in scalar QED, like the Klein paradox have no correspondence with any effect in the axion photon system. 

At the time refs. \cite{duality} and \cite{splt}
were written the Global scalar QED had not been yet been formulated, so the analogy 
of the axion photon system in an external magnetic field was made with standard scalar QED,
but we see here that the exact analogy is rather with Global scalar QED.

\section{1+1 Q-Ball like solutions in the axion-photon system}
\label{solitionsol}

We have argued that a theory with the structure similar to that of Global Scalar QED should support 
Q-Ball type solitons and in the previous section we have seen that the axion photon system in a magnetic field shares the basic features that characterize Global Scalar QED.
Now we turn therefore to consider a time dependent axion and electromagnetic field with propagation only along the $z$ direction and where a time independent magnetic field pointing in the $x$-direction is present. This field may have only a $z$ dependence, to be determined later. We want however that this field will take into account the back reaction of the time dependent axion and electromagnetic fields in a time averaged way.

Repeating shortly the manipulations of the previous sections, where we choose again to work in the temporal gauge and consider only the $x$ polarization of the electromagnetic (time dependent) fields (i.e $E_{x} = - 
\partial_{t}A$, where $A$ is the $x$ component of the vector potential), we write the interaction term as (after dropping a total time derivative)

\begin{equation}
\label{inter}
	S_{I} = -\int d^{4}x\left[gB_{x}(z)\phi E_{x}\right] = \frac{1}{2}\int dzdtgB_{x}(z)\left[\phi\partial_{t}A - A\partial_{t}\phi\right],
\end{equation}

where $B_x = -\pa_{z} A_y$, since taking also that $A_z$ depends only on $z$ leave us only with this contribution. The 1+1 dimensional effective action is

\begin{equation}
\label{ }
	S_{3} = \int dzdt\left[\frac{1}{2}\partial_{\mu}A\partial^{\mu}A + \frac{1}{2}\partial_{\mu}\phi\partial^{\mu}\phi - \frac{1}{2}m^{2}\phi^{2} + gB_{x}(z)\phi\partial_{t}A - \frac{1}{2}(\partial_{z}A_{y})^{2}\right], 
\end{equation}

so now we can discuss the eq. of motion for this magnetic mean field \cite{soliton}

\be
\pa_{z}(\frac{ig}{2}( \psi^{*}\pa_{t}\psi - \psi \pa_{t} \psi^{*}) + B_x(z)) = 0.
\lab{B eq.}
\ee

The same result can be obtained from the original equations instead of the averaged Lagrangian obtained under the assumption that the mean field $B_x(z)$ is time independent and there doing a time averaging procedure, using for example
that under such time averaging $\phi \pa_{t} A$ equals $\frac{1}{2}(\phi \pa_{t} A -\pa_{t}\phi A)$ (here, again, $A$ denotes the $x$ component of the vector potential).
Equation \rf{B eq.} can be integrated, giving

\be
\label{mean field solution}
B_x(z) = - \frac{ig}{2}( \psi^{*}\pa_{t}\psi - \psi \pa_{t} \psi^{*}) + B_0,
\ee

where $B_0$ is an integration constant. The constant $B_0$ breaks spontaneously the charge conjugation symmetry of the theory (\ref{charge conjugation}), which is equivalent to changing the sign of $A$, since in such transformation the first term in the RHS of (\ref{mean field solution}) changes sign. This would be required in order to leave the interaction term (\ref{inter}) in the action invariant. However, the second term will not change (since it is a constant). Also, in problems where $B_x$ is taken as an external field, the interaction automatically breaks this ``charge conjugation'' symmetry.

We now consider $\psi$ to have the following time dependence,

\be
\psi = \rho(z)\exp(-i \omega t).
\lab{time dep.}
\ee

We want to see now what is the equation of motion for $\rho(z)$, which we take as a real field. We begin with the general eq. for $\psi$

\be
\pa_{\mu}\pa^{\mu} \psi + igB_x(z)\pa_{0}\psi = 0.
\lab{complex eq.}
\ee

Inserting \rf{time dep.} into (\ref{mean field solution}) and the result into \rf{complex eq.}, we obtain

\be
\frac{d^{2}\rho(z)}{dz^{2}} + \frac{d V_{eff}(\rho)}{d \rho} = 0,
\lab{analog mechanical}
\ee

where $V_{eff}(\rho)$ is given by

\be
V_{eff}(\rho) = \frac{1}{2}(\omega^{2} -  \omega g B_0)\rho^{2} + \frac{1}{4}g^{2}\omega^{2}\rho^{4}.
\lab{effective potential}
\ee

Some comments are required on the nature and signs of the different terms.
One should notice first of all that this effective potential is totally dynamically generated and vanishes when taking $ \omega = 0 $. 
Concerning signs, all terms proportional to $\omega^{2}$
are positive, in fact although the $(g\omega)^{2}$ term is quartic in $\rho$, it has to be regarded as originating not from  an ordinary potential of the scalar field in the original action, but rather from a term
proportional to a  $g^{2}(- i( \psi^{*}\pa_{t}\psi - \psi \pa_{t} \psi^{*}))^{2}$, quadratic in time derivatives, which could have been obtained if we had worked directly with the action rather than with the equations of motion, replacing (\ref{mean field solution}) back into the action i.e., integrating out the $B_x$ field (generically replacing solutions back in the action does not give correct results, here the procedure  gives correct results provided a lagrange multiplier enforcing magnetic flux conservation of (\ref{mean field solution}) is added, but this does not affect the terms dicussed here). Such type of quadratic terms in the time derivatives give a positive contribution both in the lagrangian and in the energy density, unlike a standard (not of kinetic origin) potential, where the contribution to the lagrangian is opposite to that of their contribution in the energy density.

The only term which may not be positive is the $-\omega g B_0$ contribution. This term breaks the charge conjugation symmetry (\ref{charge conjugation}) which for a field of the form \rf{time dep.} means $\omega \rightarrow -\omega$.

We can in any case choose the sign of $\omega$ such that the $-\omega g B_0$ contribution is negative and choose big enough  $B_0$
(or  $\omega $ small enough) so that this term makes the first term in the effective potential negative.

Now we are interested in obtaining solutions where 
$B_x(z) \rightarrow B_0$ as $z \rightarrow  \infty$ and also as $z \rightarrow  -\infty$, which requires $\rho \rightarrow 0$ as $z \rightarrow  \infty$ and also as $z \rightarrow  -\infty$. Since the vacuum with only a constant magnetic field is a stable one \ct{Ansoldi}.

The solution of the equations \rf{analog mechanical} and \rf{effective potential} with such boundary conditions is possible if
$\omega^{2}  - \omega g B_0 < 0$. After solving these analog of the "particle in a potential problem" with zero ``energy'', so that the boundary conditions are satisfied, we find that $\rho$ is given by (up to a sign),

\be
\rho = \frac{(\sqrt{2(\omega g B_0 - \omega^{2}) })/g\omega}{cosh(\sqrt{\omega g B_0 - \omega^{2} }(z-z_0))},
\lab{the solution}
\ee

where $z_0$ is an integration constant that defines the center of the soliton.

Inserting \rf{the solution} and \rf{time dep.} into the expression for $B_x$ (\ref{mean field solution}) we find the profile for
$B_x$ as a function of $z$. The difference in flux per unit length (that is ignoring the integration with respect to $y$ in the $yz$ plane) through the $yz$ plane of this solution with respect to the background solution
$B_x=B_0$ is finite amount. Since magnetic flux is conserved, we take this as an indication of the stability of this solution towards  decaying into the $B_x=B_0$ stable "ground state". 

Notice also that the soliton is charged under the $U(1)$ axion photon duality symmetry (\ref{axion photon symm}) and the vacuum is not, another evidence for the stability of these solitons. However for any given soliton,
there is no "antisoliton", since the condition $\omega^{2} -  \omega g B_0 < 0$ will not be mantained if we reverse the sign of $\omega$.
This is due to the fact that the vacuum of the theory, i.e. $B_x=B_0$ spontaneously breaks the charge conjugation symmetry (\ref{charge conjugation}).

\section{Adding an extra scalar with Goldstone (gradient) couplings and de-localization of the 4-point interactions}

So far we have formulated the basic idea concerning the Global scalar QED by using the auxiliary field $B$. We have also discussed several situations, when there is spontaneous symmetry breaking, Q-balls solitons and the theory under an external potential, for this case it is interesting that a system of axions and photons in an external magnetic field can be reduce to this type of models.
It is also interesting that the theory under spontaneous symmetry breaking seems to more suitable to a perturbative treatment, gauge fixing and so on than the theory without spontaneous symmetry breaking.
On the other hand, the unbroken theory is able to describe really charged particles and not just Goldstone bosons. We  can combine the benefits of both the broken theory and the unbroken theory by considering
an extra real scalar with Goldstone (gradient) couplings in addition to the complex scalar field $\psi$.
 A very simple one is defined by the model that introduces the massless field $A$
 which provides the additional  $U(1)$ symmetry  $A \rightarrow A + constant$, since there is no mass associated to this field  and its couplings are derivative,
\begin{equation} \label{GlobalQED-U(1)xU(1)}
\mathcal{L}= g^{\mu\nu}\frac{\partial\psi^{*}}{\partial x^{\mu}}\frac{\partial \psi}{\partial x^{\nu}}-U(\psi^{*}\psi) -\frac{1}{4} F^{\mu \nu}F_{\mu \nu} + j_\mu(\psi) (A^{\mu} +\partial^{\mu}B)
 +\partial_\mu A (A^{\mu} +\partial^{\mu}B)
\end{equation}
 
 where
 
 \begin{equation}
 j_\mu (\psi) =  ie(\psi^{*}\frac{\partial\psi}{\partial x^\mu}  -\psi\frac{\partial\psi^{*}}{\partial x^\mu})
\end{equation}
The Maxwell's equations are now
\begin{equation}
\partial^\nu F_{\nu \mu } = j_\mu (\psi) +  \partial_\mu A
\end{equation}
 
Note that the equation obtained from the variation of  $A$ is very simple,
it gives us

\begin{equation}
  \partial^\mu(A_{\mu} +\partial_{\mu}B) = 0
\end{equation}  
 which is formally solved as
$B= -\frac{1}{\Box}\partial^\mu A_{\mu}$.

No difficulties result in taking the gauge $B=0$ now.

Also, by taking the divergence of the Maxwell's equation, i.e., from the conservation of the total current, we obtain 
\begin{equation}
A= -\frac{1}{\Box}\partial^\mu j_{\mu}(\psi)
\end{equation} 
 The conserved Noether current obtained from the independent global phase transformations
$\psi \rightarrow exp (i\chi) \psi$,  $\chi$ being a constant, is still given by

\begin{equation}
 J_\mu =  ie(\psi^{*}\frac{\partial\psi}{\partial x^\mu}  -\psi\frac{\partial\psi^{*}}{\partial x^\mu})
 + 2e^{2} (A_{\mu} +\partial_{\mu}B)\psi^{*}\psi
\end{equation}
The conservation of the Noether current allows us to obtain an expression for 
$\partial^\mu j_{\mu}(\psi)$,
 \begin{equation}
\partial^\mu j_{\mu}(\psi) =
 - 2e^{2} \partial^\mu ((A_{\mu} +\partial_{\mu}B)\psi^{*}\psi) = 
  - 2e^{2}  (A_{\mu} +\partial_{\mu}B)\partial^\mu(\psi^{*}\psi)
\end{equation}
Therefore the Maxwell's equations become now
\begin{equation}
\partial^\nu F_{\nu \mu } = j_\mu (\psi) +  \partial_\mu \frac{1}{\Box} ( 2e^{2}  (A_{\nu} +\partial_{\nu}B)\partial^\nu(\psi^{*}\psi))
\end{equation}
 
We see that in this version of Global scalar QED the sea gull term in the current of the standard scalar QED ($e^{2}$ term in eq. (\ref{KG:chargedensityandcurrent})) which is obtained from a four point interaction, is replaced by a smoothed out interaction.

A model similar to this, with the $A$ field was studied was studied in ref \cite{Guendelman} and although it does not coincide with what we deal here, it does in the case where we let the interactions go to zero. From this comparison with  ref \cite{Guendelman} we can find out that the field $A$ can be responsible for zero norm states. Unlike what was discussed in ref \cite{Guendelman}, the $A$ field becomes here an interacting field when we turn $e$ on. We discuss in the next section a way to avoid these kind of ghost states with a simple generalization, which also introduces a scale into the problem.

\section{Ghost free model with a new scale}
To avoid ghost states associated with the field $A$ we can add a "wrong sign" kinetic term for this field, but when the $B$ field is integrated out, the whole system produces a correct sign kinetic term for $A$, so we consider,
\begin{equation} \label{GlobalQED-U(1)xU(1)NG}
\mathcal{L}= g^{\mu\nu}\frac{\partial\psi^{*}}{\partial x^{\mu}}\frac{\partial \psi}{\partial x^{\nu}}-U(\psi^{*}\psi) -\frac{1}{4} F^{\mu \nu}F_{\mu \nu} + j_\mu(\psi) (A^{\mu} +\partial^{\mu}B)
 - \frac{1}{2}g^{\mu\nu}\frac{\partial A}{\partial x^{\mu}}\frac{\partial A}{\partial x^{\nu}}
  +\frac{1}{l_0}\partial_\mu A (A^{\mu} +\partial^{\mu}B)
\end{equation}
 
 where still
 
 \begin{equation}
 j_\mu (\psi) =  ie(\psi^{*}\frac{\partial\psi}{\partial x^\mu}  -\psi\frac{\partial\psi^{*}}{\partial x^\mu})
\end{equation}

Since the kinetic term determines the dimensionality of the field $A$ to be that of a canonical scalar field, we are forced now to introduce a new scale of length $l_0$, i.e. a new scale in the theory.
The Maxwell's equations are now
\begin{equation}
\partial^\nu F_{\nu \mu } = j_\mu (\psi) +  \frac{1}{l_0} \partial_\mu A
\end{equation}
 
Note that the equation obtained from the variation of  $A$ is now, it gives us

\begin{equation}\label{BAequation}
-\Box A +  \frac{1}{l_0}\partial^\mu(A_{\mu} +\partial_{\mu}B) = 0
\end{equation}  
 which is formally solved as
$B= -\frac{1}{\Box}\partial^\mu A_{\mu} + l_0 A$. Notice that when reintroducing this back into the original action, we generate a correct sign kinetic term for $A$ (i.e. this is not a ghost field).

One can understand the ghost free nature of the model  more rigorously by considering the gauge $B=l_0A$, which implies 
using  equation (\ref{BAequation}), that we obtain the Landau gauge condition $\partial^\mu A_{\mu} =0$. Inserting $l_0 A$ instead of $B$  in the action, we obtain a ghost free action, with $A$ being a normal field, not a ghost. Any ghosts remaining are related to the  Landau gauge condition  $\partial^\mu A_{\mu} =0$, and the residual gauge freedom in this gauge, but this is a well understood subject, where the ghosts do not mix in the interaction with physical particles. 
 
The Landau gauge plus its residual gauge freedom reduce, as is well known, is able to reduce the physical  photon degrees of freedom to two independent polarizations. We have also two charged particles from the $\psi$ field and the scalar particles generated by the field $A$.

 The conserved Noether current obtained from the independent global phase transformations
$\psi \rightarrow exp (i\chi) \psi$,  $\chi$ being a constant, is still given by

\begin{equation}
 J_\mu =  ie(\psi^{*}\frac{\partial\psi}{\partial x^\mu}  -\psi\frac{\partial\psi^{*}}{\partial x^\mu})
 + 2e^{2} (A_{\mu} +\partial_{\mu}B)\psi^{*}\psi
\end{equation}
The conservation of the Noether current allows us to obtain an expression for 
$\partial^\mu j_{\mu}(\psi)$,
 \begin{equation}
\partial^\mu j_{\mu}(\psi) =
 - 2e^{2} \partial^\mu ((A_{\mu} +\partial_{\mu}B)\psi^{*}\psi)  
\end{equation}
and again, by taking the divergence of the Maxwell's equation, i.e., from the conservation of the total current, we obtain 
\begin{equation}
A= -l_0 \frac{1}{\Box}\partial^\mu j_{\mu}(\psi)
\end{equation} 

This is consistent also with the variation of the action with respect to $B$.
Therefore the Maxwell's equations become now
\begin{equation}
\partial^\nu F_{\nu \mu } = j_\mu (\psi) +  \partial_\mu \frac{1}{\Box} 2e^{2} \partial^\mu ((A_{\mu} +\partial_{\mu}B)\psi^{*}\psi)  
\end{equation}
 
We see again that in this (in this slightly more complicated but ghost free) version of Global scalar QED the sea gull term in the current of the standard scalar QED ($e^{2}$ term in eq. (\ref{KG:chargedensityandcurrent})), which is obtained from a four point interaction, is replaced by a smoothed out interaction.

To get more explicit expressions we work now a bit the equation for the field $A$. We have that
$\Box A= -l_0 \partial^\mu j_{\mu}(\psi)= l_0 2e^{2} \partial^\mu ((A_{\mu} +\partial_{\mu}B)\psi^{*}\psi)$ 

$=  l_0 2e^{2} \partial^\mu (A_{\mu} +\partial_{\mu}B)\psi^{*}\psi+l_0 2e^{2}(A_{\mu} + \partial_{\mu}B) 
 \partial^{\mu}(\psi^{*}\psi)$, now using that 
 $ \partial^\mu(A_{\mu} +\partial_{\mu}B)= l_0 \Box A $ , we get
 \begin{equation}\label{Asolequation}
 \Box A = \frac{l_0 2e^{2}}{1- l_0^{2} 2e^{2}\psi^{*}\psi}(A_{\mu} + \partial_{\mu}B) 
 \partial^{\mu}(\psi^{*}\psi)
\end{equation}
Therefore the Maxwell's equations become now
\begin{equation}\label{Maxwelleq}
\partial^\nu F_{\nu \mu } = j_\mu (\psi) +  \partial_\mu \frac{1}{\Box} ( \frac{2e^{2}}{1- l_0^{2} 2e^{2}\psi^{*}\psi}(A_{\mu} + \partial_{\mu}B) 
 \partial^{\mu}(\psi^{*}\psi))
\end{equation}
 
We see that in this version of Global scalar QED the sea gull term in the current of the standard scalar QED ($e^{2}$ term in eq. (\ref{KG:chargedensityandcurrent})) which is obtained from a four point interaction, is replaced by a smoothed out interaction. In the limit $ l_0 \rightarrow 0$, we obtain the expression of the previous section.
For $ l_0 \neq 0$, integrating out the $A$ field gives rise to non polynomial interactions. 

The $A$ particle production goes like $l_0 2e^{2}$, so as $l_0 \rightarrow 0$ the  $A$ particle production
goes to zero, but the charge production from the $A$ particles does not go to zero. This is consistent with what we mentioned that the $A$ field, in the case the added kinetic term is absent, when rescaling the $A$ field so that we can obtain a smooth limit in this limit $l_0 \rightarrow 0$, which corresponds to the model of the previous section, where $A$ creates only zero norm states, so there cannot be real particle production in that case.

Notice that eq. (\ref{Asolequation}) may or may not provide  a solution for $A$, For example it provides a solution for the $B=0$ gauge, but we may be more interested in the ghost free gauge $B=l_0A$
in which case $A$ appears in both sides of the equation, but then a well defined perturbative expansion is possible.
\begin{equation}\label{AnoBsolequation}
 \Box A = \frac{l_0 2e^{2}}{1- l_0^{2} 2e^{2}\psi^{*}\psi}(A_{\mu} + l_0\partial_\mu A) 
 \partial^{\mu}(\psi^{*}\psi)
\end{equation}
which can be written as an integral equation
\begin{equation}\label{AnoBsolequation1}
  A = \frac{1}{\Box}(\frac{l_0 2e^{2}}{1- l_0^{2} 2e^{2}\psi^{*}\psi}(A_{\mu} + l_0\partial_\mu A) 
 \partial^{\mu}(\psi^{*}\psi))
\end{equation}
for which standard methods of iterating the expression of $A$ in the right hand side can be used to obtain a well defined perturbative solution in this gauge.

Before finishing this section we want to point out an interesting feature of the $B= l_0 A$ gauge, which is that indeed inserting this in the action leaves with an action which is capable of reproducing for us all the equations of motion and the gauge fixing. This is unusual, since in ordinary electrodynamics setting for example the temporal gauge and setting in the lagrangian $A_0=0$ leaves us with an action that cannot reproduce Gauss's law. In contrast to this, setting $B= l_0 A$ in the lagrangian, produces the gauge fixed action

\begin{equation} \label{GlobalQED-gaugefixedNG}
\mathcal{L}_{gauge-fixed}= g^{\mu\nu}\frac{\partial\psi^{*}}{\partial x^{\mu}}\frac{\partial \psi}{\partial x^{\nu}}-U(\psi^{*}\psi) -\frac{1}{4} F^{\mu \nu}F_{\mu \nu} + j_\mu(\psi) (A^{\mu} + l_0\partial^{\mu}A)
 + \frac{1}{2}g^{\mu\nu}\frac{\partial A}{\partial x^{\mu}}\frac{\partial A}{\partial x^{\nu}}
  +\frac{1}{l_0}\partial_\mu A A^{\mu}
\end{equation}
 
 where still
 
 \begin{equation}
 j_\mu (\psi) =  ie(\psi^{*}\frac{\partial\psi}{\partial x^\mu}  -\psi\frac{\partial\psi^{*}}{\partial x^\mu})
\end{equation}

 The variation of the gauge fixed action with respect to the gauge potential produces for us still the same Maxwell's equations (\ref{Maxwelleq}) with the same current. One can see then that the consistency of the conservation of the electric current $\Box A= -l_0 \partial^\mu j_{\mu}(\psi)$ and the equation obtained from the variation of the $A$ field 
 $\Box A= -l_0 \partial^\mu j_{\mu}(\psi)-\frac{1}{l_0}\partial_\mu A^{\mu}$, 
implies the Landau gauge condition $\partial_\mu A^{\mu}= 0 $.

Therefore all relevant information concerning the theory and the appropriate gauge fixing is contained in  (\ref{GlobalQED-gaugefixedNG}). This gauge fixed lagrangian can be quantized canonically in a straightforward way.

Notice that because of the Landau gauge condition (also refered to as Lorenz gauge condition ), in the equation for the field $A$ there is no mixing between the $A$ field and unphysical longitudinal photons, such mixing could put in question the ghost free nature of the field $A$, but fortunately it does not exist.

\section{Discussion and Conclusions}
Discussing the new global QED makes sense from both the purely theoretical point of view, since it provides a new type of viewing interactions of charged scalar particles with electromagnetism, as well as from a phenomenological point of view, since standard scalar QED contains the sea gull contributions for which apparently do not represent any known physical process in the electrodynamics of charged pions for example, so it makes sense to build a theory without such sea gulls. 

In this framework Q-Balls \cite{Q-Balls} and other non topological solitons \cite{NTPS} that owe their existence to a global $U(1)$ symmetry can be coupled to electromagnetism and could represent multiply charged particles now in search in the LHC.

We have also shown an example of an "Emergent" global scalar QED from a system of photons and axions in an external magnetic field. This is interesting because it helps us to understand phenomena in axion-photon system in terms of a scalar QED in an external field. However the exact analogy is not with the standard
scalar QED but with the Global scalar QED. For example effects that exist in Global and standard scalar QED will take place in axion photon system in an external magnetic field, with the magnetic field assuming the role of the external electric potential. So there is a corresponding effect in axion-photon system to the 
splitting of positive and negative charges, which is the splitting of axion and photon beams in  an external magnetic field. But there is no corresponding effect to the Klein paradox in the axion-photon system, as this effect  does not exist in Global scalar QED (although it does in 
standard scalar QED),this is due to the absence of quadratic terms in the external potential 
in Global Scalar QED as opposed to standard scalar QED. These quadratic terms in the external gauge potentials are what leads to the sea gulls diagrams in perturbation theory.

As an example of the expected Q-ball like solitons, we have reviwed the ones that appear in the axion-photon system. More work on Q-ball like solitons for all types of scalar QED type theories is required.

A formulation of Global scalar QED with a new field that allows for a perturbative treatment of all degrees of freedom simultaneously is formulated. In this case additional fields need to be introduced and as a result although Global scalar QED and standard scalar QED coincide to first order in $e$, the four point interaction represented by a sea gull contribution is replaced by a smoothed out contribution.

Finally, we show that these kind of models can be generalized so as to avoid ghost states. We introduce a kinetic term for  $A$ which is of  the wrong sign, but after integrating out the $B$ field the resulting 
over-all kinetic term for the $A$ field is of the correct sign. Introducing a kinetic term forces us to introduce a new scale of length $l_0$. Integrating out the $A$ field gives rise to non polynomial interactions. The appearance of non polynomial interactions after integrating out the $A$ field does not contradict the renormalizability of the model, which should be studied in detail, since procedure of integrating the $A$ field is a non perturbative procedure.
All relevant information concerning the theory and the appropriate gauge fixing is summarized in a gauge fixed lagrangian (\ref{GlobalQED-gaugefixedNG}). This gauge fixed lagrangian can be quantized canonically in a straightforward way. 

The natural step after this is done would be to calculate fundamental processes  
like Compton scattering in this theory. Notice that these processes will be somewhat more complicated to deal with than in standard scalar QED because of the richer particle content of the theory, we hope to come back to these issues in a future publication. 

Finally, we want to point out parallel developments in \cite{Roee}, concerning the use of the auxiliary field $B$ to
 consistently and gauge invariantly formulate models where the coupling constant is a non trivial function of a scalar field . In the $U(1)$ case the coupling to the gauge field can contains a term of the form
 $g(\phi)j_\mu (A^{\mu} +\partial^{\mu}B)$ where $B$ is the auxiliary field and  $j_\mu$ is the Dirac current. The scalar field $\phi$ determines the local value of the coupling of the gauge field to the 
 Dirac particle. The consistency of the equations determine the condition $\partial^{\mu}\phi j_\mu  = 0$
which implies that the Dirac current cannot have a component in the direction of the gradient of the scalar field. As a consequence, if  $\phi$ has a soliton behaviour, like defining a bubble that connects two vacuua, we obtain that the Dirac current cannot have a flux through the wall of the bubble, defining a confinement mechanism where the fermions are kept inside those bags. Consistent models with time dependent fine structure constant can be also constructed \cite{Roee}. This gives another area in which these ideas can be studied.

\section{Acknowledgements}
I am very grateful to Holger Nielsen, Anatoli Romaniouk,  Norma Susana Mankoc Borstnik and Jon Chkareuli for very useful conversations, as well as to all the participants of the conference "What is beyond the Standard Model", 16th Conference , Bled, July 14-21, 2013 for additional feedback on my presentation at this conference on the subject of this paper. I am also very grateful to Patricio Cordero and Gonzalo Palma for useful comments and for inviting me to the Facultad de Ciencias Fisicas y Matematicas of the Universidad de Chile for a seminar on this subject, also my gratitude to
 Alfonso Zerwekh, Ivan Smith , Amir Rezaeian for useful comments and for inviting me to the Universidad Federico Santa Maria in Valparaiso,  for a seminar on this subject. Finally most special thanks to the hospitality provided by the Universidad Catolica de Valparaiso, in particular to Sergio del Campo and Ramon Herrera and all the participants of a seminar there for useful comments and to the Institue for Nuclear Research and Nuclear Energy of the Bulgarian Academy of Sciences in Sofia, Ivan Todorov,  Emil Nissimov, Svetlana Pacheva, 	
Tchavdar Palev and other members of the academy for their hospitality 
 and the possibility of delivering a talk on the subject and obtaining important additional back reaction.

\end{document}